\documentclass[pre,floats,aps,amssymb,preprint,epsf,epsfig,rotate]{revtex4}
\usepackage{graphicx}
\begin{document}

\title{Dynamic properties in a family of competitive growing models}

\author{Claudio M. Horowitz and Ezequiel V. Albano.}

\affiliation{Instituto de Investigaciones Fisicoqu\'{\i}micas Te\'{o}ricas
y Aplicadas, (INIFTA), CONICET, UNLP.
Sucursal 4, Casilla de Correo 16, (1900) La Plata. ARGENTINA. FAX : 0054-221-4254642.
E-mail : horowitz@inifta.unlp.edu.ar}

\date{\today}


\begin{abstract}

The properties of a wide variety of growing models, generically called $X/RD$,  
involving the deposition of particles
according to competitive processes, such that a particle is attached to the
aggregate with probability $p$ following the mechanisms of a generic model
$X$ that provides the correlations and at random (Random Deposition (RD))
with probability $(1 -p)$, are studied by means of numerical simulations
and analytic developments. The study comprises the following $X$ models: 
Ballistic Deposition, Random Deposition with Surface Relaxation, 
Das Sarma-Tamboronea, Kim-Kosterlitz, Lai-Das Sarma, 
Wolf-Villain, Large Curvature, and three additional models 
that are variants of the Ballistic Deposition model.

It is shown that after a growing regime,
the interface width becomes saturated at a crossover time ($t_{x2}$)
that, by fixing the sample size, scales with $p$ according
to $t_{x2}(p)\propto p^{-y}, \qquad (p > 0)$, where $y$ is an
exponent. Also, the interface width at saturation 
($W_{sat}$) scales as $W_{sat}(p)\propto p^{-\delta },  \qquad (p > 0)$,   
where $\delta$ is another exponent.
 
It is proved that, in any dimension, 
the exponents $\delta$ and $y$ obey the following 
relationship: $\delta = y \beta_{RD}$,
where $\beta_{RD} = 1/2$ is the growing exponent for $RD$.
Furthermore, both exponents exhibit universality in the 
$p \rightarrow 0$ limit.

By mapping the behaviour of the average height difference of 
two neighbouring sites in discrete models of 
type $X/RD$ and two kinds of random walks, 
we have determined the exact value of the exponent $\delta$.
When the height difference between two neighbouring sites corresponds 
to a random walk that after walking $\langle n \rangle$ steps 
returns to a distance from its initial position that is proportional to
the maximum distance reached (random walk of Type A), 
one has $\delta = 1/2$. On the other hand, when the height difference between two 
neighbouring sites corresponds to a random walk that after 
$\langle n \rangle$ steps moves $\langle l \rangle$ steps towards the initial 
position (random walk of Type B), one has $\delta=1$.

Finally, by linking four well-established universality classes (namely 
Edwards-Wilkinson, Kardar-Parisi-Zhang, Linear-MBE and Non-linear-MBE)
with the properties of Type A and B of random walks, eight different
stochastic equations for all the competitive models studied
are derived.

\vskip 0.5 true cm



\end{abstract}
\maketitle

\pagebreak

\section{INTRODUCTION}

The study and understanding of the properties of growing interfaces have
attracted great interest. In fact, interfaces are ubiquitous in Nature 
and their study has opened a promising field of multidisciplinary 
research \cite{1,2,3,4}. Interfaces naturally emerge in a wide variety 
of systems such as film growth by vapour deposition, chemical deposition 
or molecular beam epitaxy \cite{1,5}, propagation of fire fronts \cite{8}, 
diffusion fronts \cite{9a}, 
bacterial growth \cite{6}, solidification \cite{61}, 
propagation of reaction fronts in catalysed reactions \cite{7}, 
electrodeposition/ dissolution experiments \cite{sha}, 
sedimentation \cite{5b}, etc. 

Models of growing
interfaces may be defined and studied either by using discrete lattices or by
means of continuous equations.
Discrete models are defined by a
set of rules that provide a detailed microscopic description of the evolution
of the growing aggregate. In
these models the interface is described by a discrete set ${h(i,t)}$ that
represents the height of site $i$ at time $t$. The interface has $L^d$ sites,
where $L$ is the linear size and $d$ is the dimensionality of the
substrate. The interface of the aggregate is characterised 
through the scaling behaviour of the interface width $W(L,t)\equiv 
\sqrt{1/L^{d}\sum_{i=1}^{L^d}[h(i,t)-\left\langle  h(t)\right\rangle ]^2} $.
For this purpose, the Family-Vicsek phenomenological scaling 
approach \cite{5b,5a} has proved to be very successful for the description of the
dynamic evolution of growing interfaces. In fact,
it may be expected that $W(L,t)$ would show the
spatio-temporal scaling behaviour given by \cite{5b,5a}: 
 $W_{sat} \propto L^{\alpha}$ for $t \gg t_{c}$ and
$W(t) \propto t^{\beta}$ for $t \ll t_{c}$, where $t_{c} \propto L^Z$ is the
crossover time between these two regimes. The scaling exponents $\alpha$,
$\beta$ and $Z= \alpha/\beta$ are called roughness, 
growth and dynamic exponents, respectively. Also, different
models can be grouped into universality classes when they share the 
same scaling exponents.   

In contrast to the microscopic details of the growing mechanisms 
of the interface, continuous equations focus on the  
macroscopic aspects of the roughness. Essentially, the aim is to  
follow the evolution of the coarse-grained height function $h({\mathbf x},t)$ 
by using a well-established phenomenological approach that takes 
all the relevant processes that survive at a coarse-grained level
into account.
This procedure normally leads to stochastic non-linear partial differential
equations that, in general, may be written as follows \cite{1,6a,7a,8a}     

\begin{equation}   \frac{\partial h({\mathbf x},t)}{\partial t} =
G_{j}\{h({\mathbf x},t)\} +  F + \eta ({\mathbf x},t),   \label{eq14} 
\end{equation} 

\noindent where the index $j$ symbolically denotes different 
processes, $G_{j}\{h({\mathbf x},t)\}$ is a local functional that contains
the various surface relaxation phenomena and only depends on the  
spatial derivatives of $h({\mathbf x},t)$ since the growth process 
is assumed to be determined by the local properties of the surface only. 
Also, $F$ denotes the mean 
deposition rate and $\eta ({\mathbf x},t)$ is the deposition noise that
determines   the fluctuations of the incoming flux around its mean
value $F$. It is usually assumed that the noise is spatially and temporally
uncorrelated.

In order to establish the correspondence between a continuous growth
equation and a discrete model one can apply at least three 
different methods: (i) to numerically simulate the model and compare 
the obtained scaling exponents with
those of the corresponding continuous equation, (ii) to develop a set of
plausibility arguments using physical principles and (iii) 
to derive the continuous equation analytically starting from
a given discrete model. 

There are few papers in the direction of the last method.
For example, a systematic approach proposed by Vvedensky et al
\cite{9}, where the continuous equations can be constructed directly from the
growth rules of some discrete models, based on the master equation description,
has been applied successfully\cite{9,10,11,12}. This procedure requires a
regularisation step, in which non-analytic quantities are expanded and
replaced by analytic approaches, e.g., the step function is approximated by
a shifted hyperbolic tangent function expanded in a Taylor series.
As pointed out by P\v redota and Kotrla \cite{11}, the choice of the
regularisation scheme for the step function is ambiguous. Thus, the
coefficients entering in the derived continuum 
stochastic equation cannot be determined
uniquely. Another method has shown the connection between the
ballistic deposition discrete model and the Kardar-Parisi-Zhang 
(KPZ) equation in $d=(1+1)$ dimensions. However, this method 
is not successful in $d=(2+1)$ dimensions \cite{13}.

It is worth mentioning that most of the already mentioned 
progress in the understanding of the properties of interfaces 
has been achieved when the growth of the aggregate is due to 
one kind of particle only.
In contrast, less attention has been drawn to the study 
of the dynamics of competitive growing processes. It is well known that these 
competitive processes are
significant to the growth of real materials in at least two different ways: 
a) when the growing process involves two or more kinds of
particles, and b) when deposition of a single kind of particle 
is considered, but such type of particle may undergo 
different growing mechanisms. 

One example of case a) arises from the deposition of 
alloys or systems with impurities, see e.g. \cite{A,B,C,D,H,IJ} and
references therein. In this case, there may be different interactions 
between different kinds of particles causing the growing 
mechanisms to change \cite{A,B,C,D,H,IJ}. Based on these ideas,
Cerdeira et al \cite{A,B,C,D} have studied various models
for binary systems involving competitive random-like 
and ballistic-like  deposition. 
Recently, the scaling behaviour of a two-component
surface-growth model has been studied by  Kotrla et al \cite{IJ}.
This study addresses the relationship between kinetic roughening
and phase ordering in a $(1 + 1)$-dimensional single-step
solid-on-solid model with Ising-like interactions between 
two components. 

On the other hand, considering the deposition of one kind of 
particle (case b), Pellegrini et al \cite{E,F} have studied
a ballistic model of surface growth that considers
``sticky'' and ``sliding''particles. The model
interpolates between a standard ballistic model 
when only sticky particles are deposited (with 
probability $P = 1$) and a completely restructured 
ballistic model for $P = 0$ when only unrestricted
sliding particles are allowed to become attached to the 
sample. Using this model Pellegrini et al \cite{E,F} have
given evidence of a roughening transition in dimensions 
$d = 3$ and $d = 4$, while such kind of transition is no longer
observed in $d = 2$. 

In a related context of
competitive growing processes, we have also studied two competitive growth 
models in $(1+1)$, $(2+1)$, and $(3+1)$ dimensions \cite{we1,we2,we3}.
In the first discrete growth model, namely the RDSR/RD model, the
same types of particles are aggregated according to the rules of
random deposition with
surface relaxation (RDSR) with probability $p$ and
according to the rules of random deposition (RD) with probability
($1-p$)\cite{we1}. In the second discrete growth model, namely the BD/RD
model, particles  are aggregated according to the rules of ballistic
deposition (BD) with probability $p$  and according to the rules of  
random deposition (RD) with probability ($1-p$)\cite{we2}. 

For both the RDSR/RD and the BD/RD models 
the saturation process of the interface width depends
sensitively on $p$:  saturation takes place at longer times for
smaller values of $p$, while the final width of the interface is 
smaller for larger $p$-values.  Furthermore, in both models, 
three different regimes and two corresponding  
crossovers can easily be observed.  
For short times, say $t < t_{x1}$, the random growth of the interface is  
observed (i.e. the RD process dominates). At this stage, correlations  
have not been developed yet 
and $W(t)$ $\propto t^{\beta _{RD}}$ ($t < t_{x1}, \beta_{RD} = 1/2$) holds.  
During an intermediate 
time regime, say $t_{x1} < t < t_{x2}$, correlations develop 
since the RDSR (BD) process now dominates leading to $W(t)$ $\propto t^{\beta
_{RDSR}}$ ($W(t)$ $\propto t^{\beta
_{BD}}$).  At a later stage, for $t > t_{x2}$, correlations can no
longer grow due to the geometrical constraint of the lattice size and
saturation is observed.    
The saturation value of the interface width  
$W_{sat}(L,p)$ and the characteristic crossover time $t_{x2}$ behave as
\cite{we1,we2,we3}  

\begin{equation} 
W_{sat}(L,p)\propto L^{\alpha _{X}}\, p^{-\delta }  \qquad (p > 0),  
\label{eq7} 
\end{equation} 

\noindent and 
 
\begin{equation} 
t_{x2}(L,p)\propto L^{Z_{X}}\,p^{-y} \qquad (p > 0),  
\label{eq8} 
\end{equation} 

\noindent respectively. Here, $\delta$ and $y$ are exponents 
and X $\equiv$ RDSR or BD, depending on the model. On the other hand,
one has that the crossover time $t_{x1}$ also scales with $p$ as $t_{x2}$ does
(see equation (\ref{eq8})).

In these previous studies we have shown that the exponents $y$ and $\delta$ 
are independent of the dimensionality. For the RDSR/RD (BD/RD) model
we have found that $\delta \approx 1 (\approx 1/2)$ 
and $y \approx 2 (\approx 1)$ \cite{we1,we2,we3}. Based on these numerical
estimates we have conjectured the following exact values
$\delta = 1$ and $y = 2$ for RDSR/RD, and  
$\delta = 1/2$ and $y = 1$ for BD/RD. Very recently, this early conjecture
has proved to be correct by using an exact analysis \cite{lidia}. 
Furthermore, these values allowed us to formulate another 
conjecture by statintg that $\delta = y / 2$ for both models \cite{we1,we2,we3}.

Also, we have shown that the stochastic representation of the RDSR/RD model
is given by \cite{we1,we2,we3} 

\begin{equation}  
\frac{\partial h({\mathbf x},t)}{\partial t}=
F + \nu_{o} p^{2}\nabla ^{2}h({\mathbf x},t) + \eta({\mathbf x},t),  
\label{eq22} 
\end{equation}

\noindent while for the BD/RD model one has

\begin{equation}  
\frac{\partial h({\mathbf x},t)}{\partial t}=F+\nu_{o} p \nabla ^{2}h({\mathbf
x},t)+  \frac {\lambda p^{3/2}}{2}[\nabla h({\mathbf x},t)]^{2}+\eta ({\mathbf
x},t),  \label{eq31} 
\end{equation}     

\noindent where $\nu_{o}$ plays the role of an effective surface tension, 
$\lambda$ represents the lateral growth, and
$\eta ({\mathbf x},t)$ is the uncorrelated white-noise term.

Within this context, the aim of this work is to perform a systematic 
study of a wide variety of competitive growth models of the type
X/RD, with  X = DST (Das Sarma-Tamboronea) \cite{tamboronea}, 
KK (Kim-Kosterlitz) \cite{KK}, LDS (Lai-Das Sarma) 
\cite{LDS}, WV (Wolf-Villain) \cite{wolf-villain}, 
LC (Large Curvature) \cite{LC}, RDSR \cite{5b}, BD \cite{1}, 
BD1, BD2 and BD3. The last three ballistic deposition-motivated models
are variants of the BD model that will be described in detail below.   
The study is focused on the behaviour of the 
exponents $y$ and $\delta$ as well as on the derivation of the 
stochastic equations of competitive models. 
For this purpose the manuscript is 
organized as follows: in section II and III
the relevant properties of $y$ and $\delta$ are addressed and the 
exact relationship between both exponents is derived, respectively. 
Subsequently, in section IV numerical results obtained by means of Monte
Carlo computer simulations covering all competitive 
models listed above are presented. 
After that, in section V, the exact values for the 
exponents $y$ and $\delta$ are obtained for all the studied models.
Section VI is devoted to the derivation of the stochastic equations 
for different competitive models while 
our conclusions are stated in section VII.\\

\section{The Relationship between $\delta$ and $y$}

In previous papers \cite{we1,we2,we3} we have shown  that for the 
RDSR/RD and the BD/RD models, the values of the exponents $\delta$ and $y$
are independent of the dimensionality of the substrate. Furthermore,
our study leads us to conjecture a simple relationship between them,
namely $\delta = y / 2$. Furthermore, we have also proposed 
and numerically tested the following phenomenological 
dynamic scaling Ansatz for both the RDRS/RD and the 
BD/RD models \cite{we1,we2,we3}

\begin{equation}
W(t,L,p) \propto L^{\alpha _{X}}p^{-\delta }F(\frac{t}{L^{Z_{X}}p^{-y}}
) ,\qquad p > 0\qquad ,t > t_{x1}\qquad ,L\rightarrow \infty \qquad,
\label{ecu.anzats0}
\end{equation}

\noindent where  X $\equiv$ RDSR or BD depending on the model and
$F$ is a suitable scaling function.
Now, if we restrict the previous Anzats by
considering variations of $p$ only (i.e. fixing $L$), 
equation (\ref{ecu.anzats0}) can be written as
 
\begin{equation}
W(t,p) \propto p^{-\delta} F^{*}(t/p^{-y}) ,\qquad p > 0\qquad ,t > 
0\qquad ,L\rightarrow \infty ,
\label{ecu.anzats}
\end{equation}

\noindent where $F^{*}(u)$ is a suitable scaling function such that:
(i) $F^{*}(u) \propto u^{\beta_{_{RD}}}$ for $u \rightarrow 0$,
(ii) $F^{*}(u) \propto u^{\beta _{X}}$  for 
$u$ in the intermediate regimen, and (iii) $F^{*}(u) = constant$ for $u \gg 1$. 
It is worth mentioning that the Anzats given by equation (\ref{ecu.anzats}) 
is valid for the three regimens ($t > 0$) while the previous one,
given by equation (\ref{ecu.anzats0}), only holds for 
regimes (ii) and (iii) with $t > t_{x1}$. This effect is due to 
the fact that by fixing the lattice size one also fixes the crossover
time $t_{x1}$ that depends on $L$.  
In order to check the validity of the new Anzats, figure 1 
shows log-log plots of $W(t,L,p)p^{\delta }$ versus 
$t/p^{-y}$ as obtained in $d=1$ dimensions, by using
values of $p$ within the range $0.01 \leq p \leq 0.64$, and for both models. 
Here, we have taken $L=256$ and $L=512$ for the RDSR/RD and the BD/RD models,
respectively.

\begin{figure}
\centerline{
\includegraphics[width=15cm,height=20cm,angle=0]{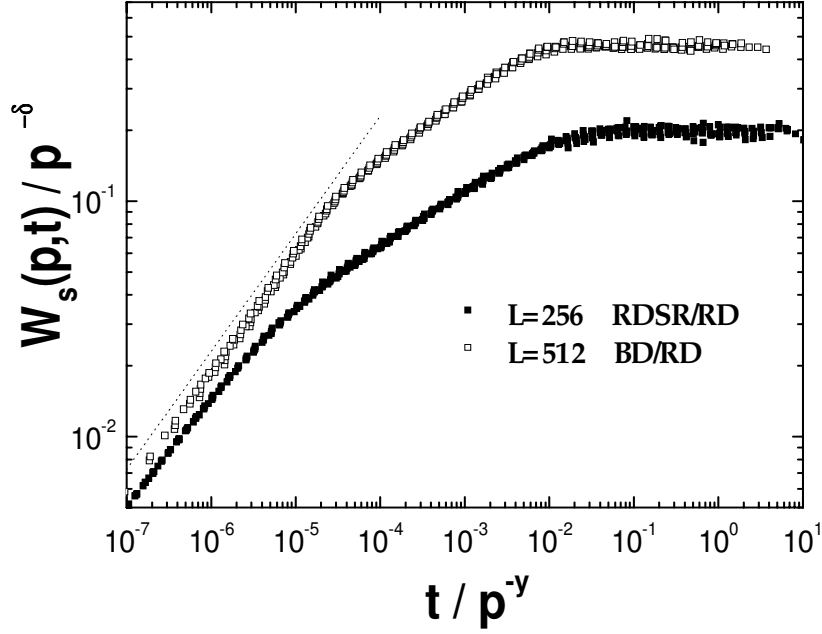}}
\caption{ Log-log plot of $W(t,L,p)/p^{-\delta }$ 
versus $t/p^{-y}$ where X = RDRS, BD. Results 
obtained for different values of $p$ ($0.01 \leq p \leq 0.64$) and lattices
 of size $L = 256$ for X=RDRS and $L = 512$ for X=BD. The data corresponding to the 
BD/RD model have been shifted one decade to the left for the sake of clarity.
The dotted line has slope $\beta_{RD} = 1/2$ and has been drawn for the
sake of comparison. More details in the text.}
\label{Fig1} 
\end{figure}

Considering the short-time regime $t < t_{x1}$, or
equivalently $u \rightarrow 0$ in equation (\ref{ecu.anzats}), we 
observed that the initial slope is independent of the considered model
and it is given by $\beta_{RD} = 1/2$, since the RD process
dominates the early stages of growth. Consequently, within the short-time
regime, equation (\ref{ecu.anzats}) can be written as     

\begin{equation}
W(t,p) \propto p^{-\delta }(t/p^{-y})^{\beta_{_{RD}}} ,
\qquad p > 0\qquad ,t < t_{x1}\qquad ,L\rightarrow \infty,
\label{ecu.anzats2}
\end{equation}

\noindent and, according to the results shown in figure 1,  
this relation has to be independent of $p$. 
So, this is true only if 

\begin{equation}
\delta =y \beta_{_ {RD}}.
\label{dely}
\end{equation}

This result strongly suggests that the factor $1/2$ already found 
in the relationship between $\delta$ and $y$ \cite{we1,we2,we3}
is just $\beta_{_ {RD}}$. It is also worth mentioning that 
$\beta_{RD}$ is independent of the dimensionality of the substrate, 
and therefore one should expect that equation (\ref{dely}) 
would also hold in any dimension.

\section{Numerical evidence on negligible finite-size corrections
to the values of the exponent $\delta$.}

\begin{figure}
\centerline{
\includegraphics[width=20cm,height=15cm,angle=0]{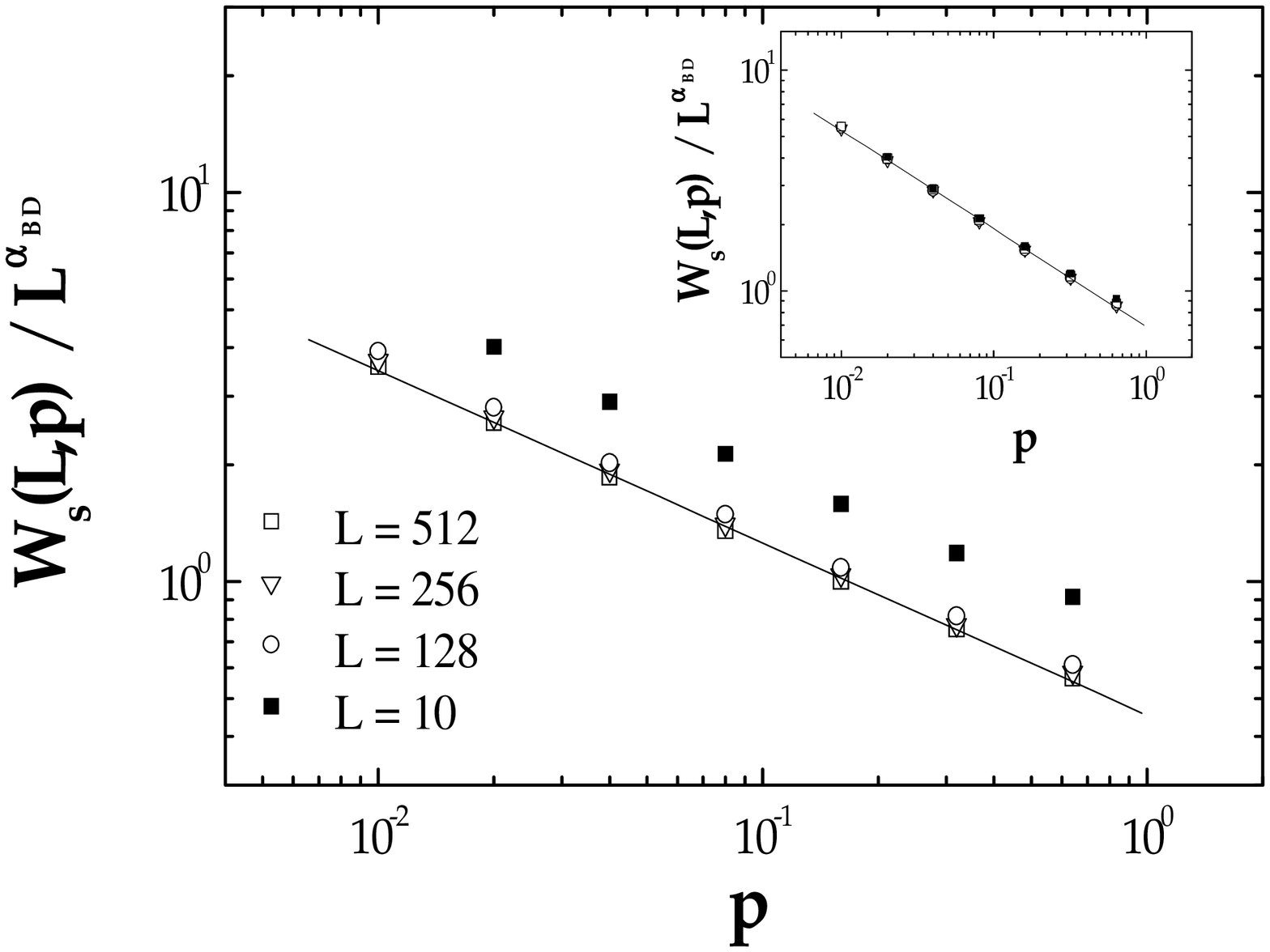}}
\caption{ Log-log plots of $W_{sat}(L,p)L^{-\alpha_{BD}}$ versus $p$
obtained for lattices of  different size, as indicated in the figure, and
assuming $\alpha_{BD} = 1/2$. The full line has slope $\delta = 0.45$
and corresponds to the best fit of the data. The insert shows the
same scaled plot but obtained assuming $\alpha_{BD}^{'} = 0.43$ for 
$L=512, 256, 128$ and  $\alpha _{BD}^{\prime }=0.46 \pm 0.05$ for 
$L=10$.
Again, the full line with slope  $\delta = 0.45$
corresponds to the best fit of the data.}
\label{Fig2} 
\end{figure}

Usually, it is observed that the numerical values of the scaling exponents 
($\alpha$, $\beta$ and $Z$) undergo systematic deviations  
when the size of the lattices used in the simulations is changed. 
Figure 2 shows log-log plots of $W_{sat}(L,p)/L^{\alpha _{BD}}$ 
versus $p$ as obtained for lattices of different size and 
taking the BD/RD model. In this figure, we have added the data 
corresponding to $L=10$ to the results already shown in paper \cite{we2}.   
Using the exact value $\alpha _{BD}=\frac{1}{2}$,
straight lines are observed, in agreement with
equation (\ref{eq7}), and the best fit of the data gives 
the slope $\delta \cong 0.45\pm 0.01$.
However, a rather small systematic deviation of the data, according to the
size of the lattice, is observed: the larger the lattice, the smaller the
ordinate. This behaviour is due to high-order corrections to scaling
that we have neglected in equation (\ref{eq7}). 
On the other hand, using the roughness
exponent obtained by fitting our data, namely 
$\alpha _{BD}^{\prime }=0.43\pm 0.05$ for 
$L=512, 256, 128$, and $\alpha _{BD}^{\prime }=0.46 \pm 0.05$ for 
$L=10$, it is possible to achieve an excellent data collapsing, as shown in
the inset of figure 2. In this case, the slope obtained by means of
a least-squares fit is also $\delta
\cong 0.45\pm 0.01$. Summing up, all results shown in figure 2
point out that the scaling Anzats given by equation (\ref{eq7}) holds
for the BD/RD model. Furthermore, we would like to emphasise that the
systematic shift of the data observed in figure 2 does not affect the 
slope of the power law, and consequently the exponent 
$\delta$ is almost independent of the lattice 
size (up to $L = 10$ in figure 2).

\begin{figure}
\centerline{
\includegraphics[width=12cm,height=8cm,angle=0]{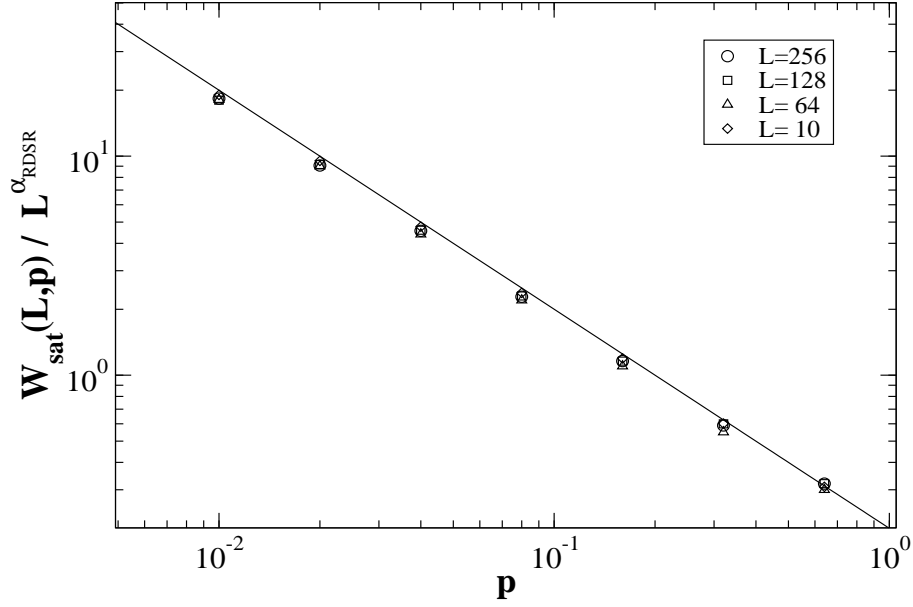}}
\caption{Log-log plots of $W_{sat}(L,p)/L^{\alpha_{RDSR}}$ versus $p$
obtained for lattices of  different size, as indicated in the figure, and
assuming $\alpha_{RDSR} = 1/2$. The full line has slope $\delta = 0.97$
and corresponds to the best fit of the data.}
\label{Fig3} 
\end{figure}

Figure 3 shows log-log plots of $W_{sat}(L,p)/L^{\alpha _{RDSR}}$ 
versus $p$ as obtained for the RDSR/RD model using 
lattices of different size. 
In this figure, we have also added the data corresponding to $L=10$ to the 
results already shown in paper \cite{we1}.   
Using the exact value $\alpha _{RDSR}=\frac{1}{2}$ for $L=512, 256, 128$ 
and $\alpha _{RDSR}=0.46$ for $L=50$,
straight lines are observed, in agreement with
equation (\ref{eq7}), and the best fit gives the slope 
$\delta \cong 0.97\pm 0.01$.

So, these results show that the values of $\delta$ are not  
appreciably affected by systematic deviations even when  
extremely small lattices are used (up to $L=10$ in the 
examples of figures 2 and 3).
It is worth mentioning that for both models we have arrived at 
the same conclusion in higher dimensions. More explicitly, we have 
performed numerical simulations by using 
$L \times L = 6 \times 6$ in $d=(2+1)$ and 
$L \times L \times L = 6 \times 6 \times 6$ in $d=(3+1)$ and we have found
$\delta \sim 0.45$ and $\delta \sim 0.97$ independently of the 
dimensionality, for the BD/RD and RDSR/RD models, respectively.
(These results are not shown here for the sake of space).

The lack of appreciable finite-size effects in the values of $\delta$ 
(up to $L=10$) is a rather surprising result.
It should be noticed that a similar behaviour is also  
found in the exponent $\beta_{RD}$ 
for the random deposition model, and it is  due to the lack of correlations  
between different columns of the aggregate.      
So, although we have not a convincing explanation for the behaviour of $\delta$ 
in our case, we think that it may be related in a non-trivial way to 
the decorrelation induced by the fraction $(1 - p)$ of particles that
are deposited according to the random deposition rules, which are 
expected to play a relevant role precisely for $p \rightarrow 0$. 
 
\section{Systematic evaluation of the exponent $\delta$ for 
a family of competitive models.}
  
In this section we take advantage of the almost negligible dependence of 
the values of the exponent $\delta$ on the lattice size, as discussed in 
the previous section, in order to evaluate it by using small lattices 
for a wide family of competitive models called X/RD. 
This family of models is defined such that particles of the same type 
are aggregated according to the rules of a generic discrete model X 
with probability $p$ and according to the rules of RD with probability
($1-p$).  

We expect that for this family of models, the dynamics of the RD process 
would play the same role as in the cases of the RDSR/RD and 
the BD/RD models. That is, RD causes the slowing down of correlations
among particles. So, we also expect that equations (\ref{eq7}) and (\ref{eq8})
would be valid, but the values of the exponents $\alpha$ and $Z$ entering in 
the scaling relationships have to be those of the X model that introduces the 
correlations among particles. Furthermore, we expect that the 
general relationship  given by equation (\ref{dely}) will also hold,
allowing us to evaluate the exponent $y$ after the determination of 
$\delta$.        


\begin{figure}
\centerline{
\includegraphics[width=10cm,height=12cm,angle=-90]{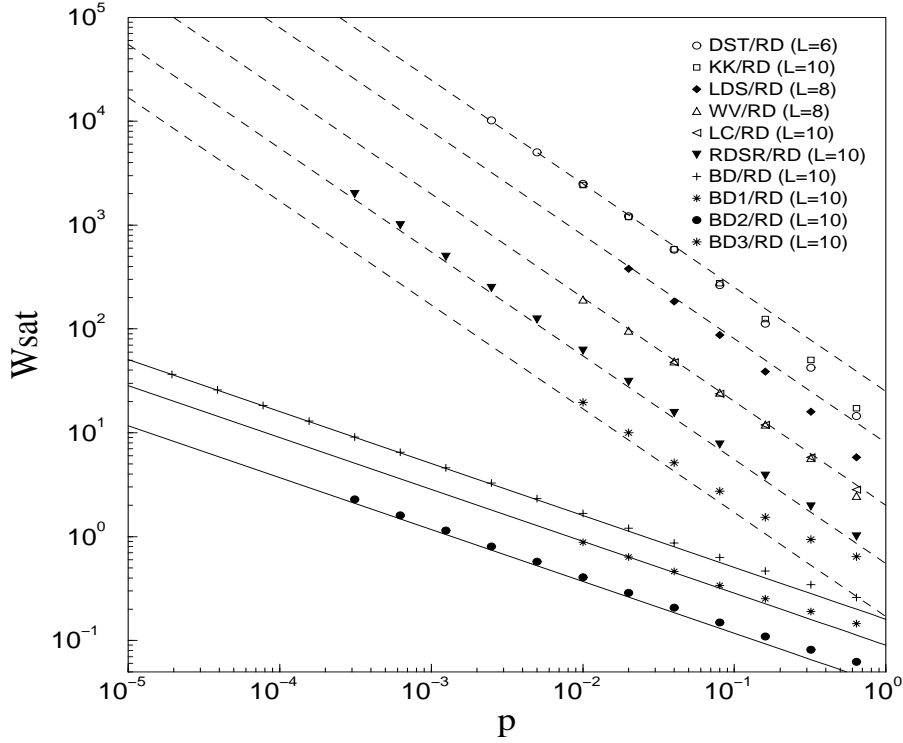}}
\caption{ Log-log plot of $W_{sat}$ versus $p$ for the following 
competitive models in dimension $d=1$: DST/RD for $L=6$,  
KK/RD for $L=10$, LDS/RD for $L=8$, WV/RD for $L=8$, 
LC/RD for $L=10$, RDSR/RD for $L=10$ , BD/RD for $L=10$ , 
BD1 for $L=10$, BD2 for $L=10$, BD3 for $L=10$.
Different plots have been shifted vertically for the sake of clarity.}
\label{Fig4} 
\end{figure}
 
Figure 4 shows log-log plots of $W_{sat}$ versus $p$, in $d=1$, as obtained 
for the family of models of the type X/RD where X = DST 
(Das Sarma-Tamboronea) \cite{tamboronea}, 
KK (Kim-Kosterlitz) \cite{KK}, LDS (Lai-Das Sarma) 
\cite{LDS}, WV (Wolf-Villain) \cite{wolf-villain}, 
LC (Large Curvature) \cite{LC}, RDSR \cite{5b}, BD \cite{1}, 
BD1, BD2, BD3. The last three models are variants of the  BD model and 
the difference between them is due to the rules used for the sticking of 
the particles that are shown schematically in figure 5. 
In fact, for the BD1 model, when the height of a selected site for the 
deposition of a particle is lower than that of a neighbouring site, 
the particle becomes stuck at half height between the selected site and 
the highest neighbouring site. 
For the case of the BD2 model, when the height of the site selected to 
deposit a particle is lower than that of a neighbouring site,
the particle becomes stuck to the highest 
neighbouring site but the actual deposition height is decreased 
by one lattice unit. Finally, for the BD3 model, when the selected site 
for the deposition of a particle is lower 
than a neighbouring site, the particle sticks leaving a single hole in 
the selected column, so after deposition the height of the selected 
site becomes enlarged by two lattice units. (See also figure 5).   

\begin{figure}
\centerline{
\includegraphics[width=10cm,height=5cm,angle=0]{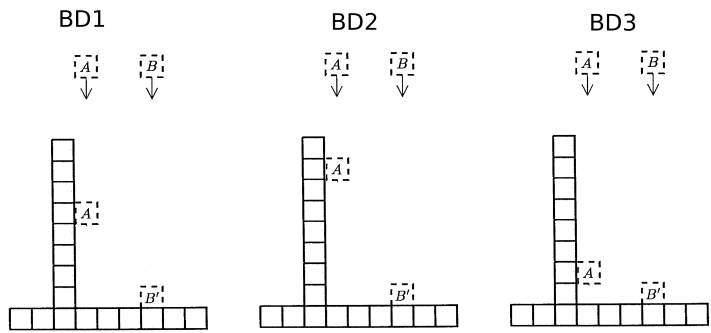}}
\caption{Schematic view of the deposition of particles in the BD1, 
BD2 and BD3 models. More details in the text.}
\label{Fig5} 
\end{figure}

As shown in figure 4, the behaviour of $\delta$ exhibits two relevant 
features. On the one hand, the obtained values of $\delta$ are independent 
of the universality class of the competitive model that introduces 
the correlations to the aggregate. So, models belonging to different 
universality classes may have the same value of $\delta$, while models 
belonging to the same universality class may have different values 
of $\delta$. On the other hand, the exponents $\delta$ obtained 
in the $p\rightarrow 0$ limit can only assume two possible values that
are very close to $\delta=1/2$ and $\delta=1$.
Further support to this statement is given in figure 6 that shows 
plots of the effective value of the exponent $\delta$ obtained 
for the family of X/RD models. 
So, we conjectured that in the $p\rightarrow 0$ limit the exact values for 
$\delta $ should be $1/2$ or $1$, depending on the model. Furthermore,
due to the large number of models considered, it would not be 
surprising that the exponent $\delta$ for all models of the type X/RD
may assume one of the already found values.  

\begin{figure}
\centerline{
\includegraphics[width=10cm,height=12cm,angle=-90]{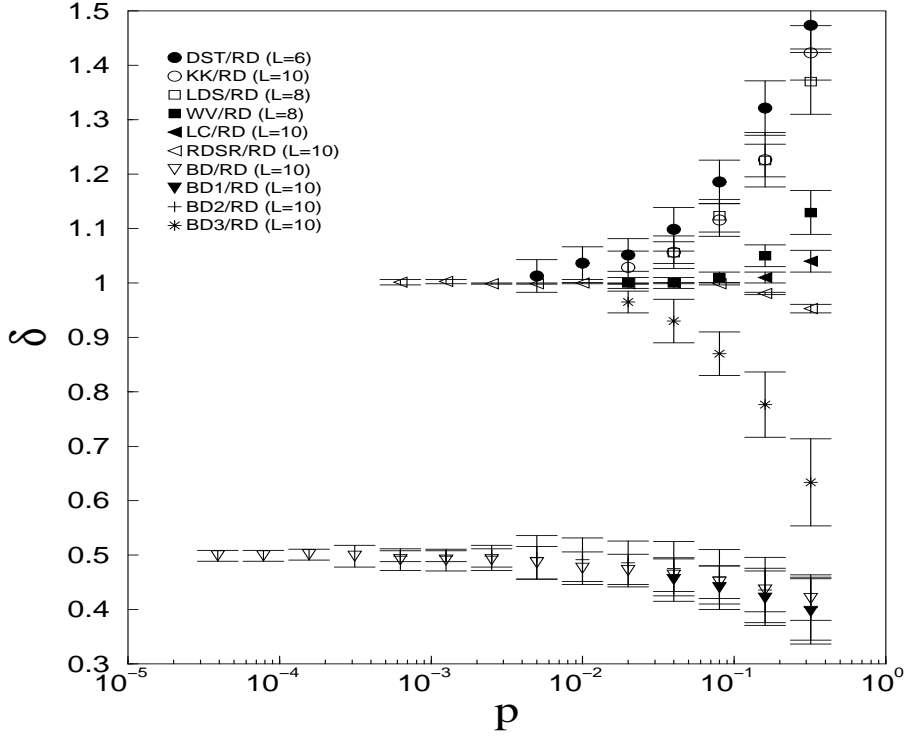}}
\caption{Log-linear plot of $\delta$ versus $p$ for the following 
competitive models in dimension $d=1$: DST/RD for $L=6$,  
KK/RD for $L=10$, LDS/RD for $L=8$, WV/RD for $L=8$, 
LC/RD for $L=10$, RDSR/RD for $L=10$, BD/RD for $L=10$ , 
BD1 for $L=10$, BD2 for $L=10$, and BD3 for $L=10$.}
\label{Fig6} 
\end{figure}

\section{Analytical calculation of the exact values of $\delta$ 
in the $p\rightarrow 0$ limit.}

In order to evaluate the exact values of $\delta$ we have used 
three concepts linking the interface width ($W$) to the height difference  
between two neighbouring sites ($h(i)-h(i+1)$) in a competitive 
growth model.

First, we take into account that from the statistical point of view, 
it is easy to show that if the interface width stops 
its growth (i.e. by reaching a saturation value), then the height difference  
between two neighbouring sites must stop its growth too. 
Secondly, we considered that if the growing stage of the interface width can be
described by a power law of the type $W(t) \propto t^{b}$, then one also 
has that $\mid h(i,t)-h(i+1,t) \mid \propto t^{b}$. Finally,  
if the dependence of the interface width on $p$ is given by another
power law, namely $W(p) \propto p^{b}$, then 
the relationship $\mid h(i,p)-h(i+1,p) \mid \propto p^{b}$ must also hold.
It is worth mentioning that the inverse relationships are also valid
in all cases. 

So, based on these statements we studied the behaviour of  
$\mid h(i)-h(i+1) \mid$ in order to obtain the relevant properties of 
the interface width.  
The simplest example for the application of the above-discussed concepts 
is provided by the RD model. Here, the height difference  
between two neighbouring sites due to the deposition of a single particle 
can only be increased or reduced by one lattice unit. 
So, this height difference corresponds to the displacement of a 
typical random walk. This fact allows us to conclude that if 
$h(i,t=0)-h(i+1,t=0) =0$ (as usual), then 
$\langle (h(i,t)-h(i+1,t))^{2} \rangle \propto t$ and 
using this result we conclude that $W(t) \propto t^{1/2}$, in agreement
with the fact that $\beta_{RD} = 1/2$.

Considering a competitive growth model, let us call ``RD particles''
and ``X particles'' those particles that are deposited according 
to RD rules and X rules, respectively. Since in the $p\rightarrow 0$ limit
the deposition probability of RD particles is larger 
than that of X particles, the deposition process in an X/RD model can 
be thought as cycles involving the deposition of $n_{i}$ RD particles followed by
a single X particle ($n_{i}$ is the number or RD particles in the cycle $i$).
Also, it is clear that from the statistical point of view, the evolution  
through this kind of cycles is equivalent to the evolution of any 
hypothetic model such that the same number $\langle n \rangle$ of 
RD particles followed by an X particle is deposited during each cycle,
where  $\langle n \rangle$ are the average number of RD particles given by 

\begin{equation}\label{intervalo1}
 \langle n \rangle  = \sum_{n=0} ^{\infty} n  (1-p)^{n}p =\frac{1-p}{p}.
\end{equation} 

Since the statistical evolution of the 
height difference $\langle h(i)-h(i+1) \rangle$
is related to the saturation width of the interface according to

\begin{equation}\label{hp}
W_{sat}(p) \propto \sqrt (\langle (h(i)-h(i+1))^{2} \rangle ),
\end{equation} 
 
\noindent our aim is to show that the behaviour of 
$W_{sat}(p)$ can also be derived from the statistical 
knowledge of $(h(i)-h(i+1))$ as a function of the deposition cycles.

On the one hand, it is straightforward to show that for models 
such as BD/RD,BD1/RD and BD2/RD, the behaviour of $h(i)-h(i+1)$ 
in each cycle corresponds to a random walk that after $\langle n \rangle$ 
steps returns either (i) to its initial position, 
(ii) to a point placed at a distance that is just half of the 
maximum distance reached from its initial position, 
or (iii) to a neighbouring site of its initial position.
So, these types of competitive models $\langle h(i)-h(i+1) \rangle $ 
correspond to a random walk that in each cycle walks 
$\langle n \rangle$ steps and after that returns to a site 
placed at a certain distance (from the starting point)  
that is proportional to the maximum distance reached 
from its initial position, $a$ being the proportionality constant.
Hereafter, this kind of random walk is called 
``Type A''. Now, when the number of cycles is large enough 
the saturation of the interface width is expected to occur
and consequently one has

\begin{equation}\label{intervalo8}
\langle (h(i)-h(i+1))^{2}  \rangle = 
\langle n \rangle \sum_{j=1} ^{\infty} a^{j} =
 \langle n \rangle \frac{a}{1-a}, 
\label{hh22}
\end{equation}

\noindent where $j$ is the number of cycles. So, using 
equation (\ref{intervalo1}) it follows 
that equation (\ref{intervalo8}) is equivalent to 
$\langle (h(i)-h(i+1))^{2} \rangle \propto 1/p$ ($p \rightarrow 0$),
so that equation (\ref{hp}) gives 

\begin{equation}\label{intervalo9}
W(t) \propto {p}^{-1/2}.
\end{equation} 

On the other hand, it is easy to show that for the BD3/RD, TDS/RD, 
LDS/RD, KK/RD, WV/RD, LC/RD and RDSR/RD models, the behaviour 
of $\langle h(i)-h(i+1) \rangle$ in each cycle corresponds to a random walk 
that after $\langle n \rangle$ steps returns
to a site placed at a distance $\langle l \rangle$ from 
the maximum distance reached
during the walk (we call this random walk ``Type B''). 
So, after the first cycle one has 

\begin{equation}\label{intervalo10}
\langle ((h(i)-h(i+1))_{1})^{2} \rangle = (\sqrt {\langle n \rangle} -l)^{2},
\end{equation}

\noindent then after the second cycle it follows that

\begin{equation}\label{intervalo11}
\langle ((h(i)-h(i+1))_{2})^{2}\rangle = (\sqrt {\langle n \rangle + 
(\sqrt {\langle n \rangle} -l)^{2}} -l)^{2},
\end{equation}

\noindent and so on. The analytic form 
of $\langle (h(i)-h(i+1))^{2} \rangle$ when the number of cycles 
is large enough can be obtained by following a simple procedure: 
Since after reaching saturation $\langle h(i)-h(i+1) \rangle$ stops growing, 
in the next cycle after saturation the increment of $\langle h(i)-h(i+1) \rangle$ 
due to the deposition of $\langle n \rangle$ RD particles 
is equal to the effect caused by the subsequent X particle.
As we have already discussed, the effect of the additional 
X particle is just to move towards 
its initial point $l$ step. So, if before the deposition of the new 
particle one had $\langle (h(i)-h(i+1))^{2} \rangle = c^{2}$ 
(where $c$ is a constant),
then after that deposition one already has 
$\langle (h(i)-h(i+1))^{2} \rangle = (c-l)^{2}$.
Then, the difference between them is given by  
$\langle (h(i)-h(i+1))^{2})_{before}\rangle -
\langle (h(i)-h(i+1))^{2})_{after} \rangle \sim 2 l \langle (h(i)-h(i+1)) \rangle$, 
and it has to be equal to the effect of the $\langle n \rangle$ RD particles.
So,

\begin{equation}\label{intervalo15}
\langle (h(i)-h(i+1))\rangle  \propto 
\langle n \rangle,
\end{equation} 

\noindent and using equation (\ref{hp}) one obtains

\begin{equation}
\label{intervalo16}
W_{sat}(p) \propto p^{-1}.
\end{equation} 
  
Therefore, we conclude that for all models that can be mapped into 
random walks of Type A and B one has that $\delta = 1/2$ 
(see equation (\ref{intervalo9})) and $\delta = 1$
(see equation (\ref{intervalo16})), respectively. 
These results provide an independent, more general,
confirmation of the exact values of the exponents 
obtained very recently for the RDSR/RD and BD/RD models \cite{lidia}.
Of course, by using equation (\ref{dely}) one can also obtain the 
exact value of the exponent $y$ for any model that can be mapped into
the two types of random walks already considered.

\section{Phenomenological Stochastic Growth Equations.}
 
It is well known that the stochastic growth equations describing 
the BD/RD and the RDSR/RD models are the KPZ 
(Kardar-Parisi-Zhang, see equation (\ref{eq31}))
and the EW (Edwards-Wilkinson, see equation (\ref{eq22}) ) 
equations, respectively. So, both models belong to the 
same universality class as that of the model that introduces the
correlations among particles, namely the X model with $X = BD$ 
and $X = RDSR$ \cite{we1,we2,we3,lidia}. 
Also, using scaling arguments on $p$ and the values of $\delta$ 
and $y$ we have found that for the BD/RD model the parameter $p$ 
appears in the linear and non-linear terms of the stochastic equation,
taking the form $\nu(p)=\nu p$ and $\lambda(p)=\lambda p^{3/2}$
(see equation (\ref{eq31})), while for the RDSR/RD model 
the parameter $p$ appears as a factor of the
form $\nu(p)=\nu p^2$ (see equation (\ref{eq22}))\cite{we1,we2,we3,lidia}.

The scaling argument on $p$ implies that, assuming the 
interface $h({\mathbf x},p,t)$ to be self-similar,
on rescaling the coordinate $p$ according to

\begin{equation}  
p\rightarrow p^{'} \equiv cp,  \label{eq31bis} 
\end{equation} 

\noindent and the height according to

\begin{equation}  
h\rightarrow h^{'} \equiv c^{-\delta}h,  \label{eq32} 
\end{equation} 

\noindent one should obtain an interface that is statistically 
indistinguishable from the original one. Since the interface 
roughness depends on time $t$ as well, one should have  

\begin{equation}  
t\rightarrow t^{'} \equiv c^{-y}t.  \label{eq33} 
\end{equation}

So, for any competitive model belonging to the X/RD family we can propose a  
stochastic growth equation similar to its X model equation where 
the parameter $p$ appears in the linear and non-linear terms.
After that, by using the scaling arguments on $p$ and the exact values of 
$\delta$ and $y$, it is possible to obtain the exact dependence of the
prefactors on $p$ for any stochastic equation. This systematic procedure and
the results of the previous section allows us to conclude that 
the behaviour of $(h(i)-h(i+1))$ quantitatively determines the 
exact dependence of the stochastic equation on $p$ .

Summing up, by using this procedure we have found that
in the cases of BD1/RD, BD2/RD and BD/RD models, 
the KPZ stochastic growth equation is given by

\begin{equation}  
\frac{\partial h({\mathbf x},t)}{\partial t}=F+\nu p \nabla ^{2}h({\mathbf
x},t)+  \frac {\lambda p^{3/2}}{2}[\nabla h({\mathbf x},t)]^{2}+\eta ({\mathbf
x},t).  \label{estocastica0} 
\end{equation}     
 
On the other hand, for the KK/RD and BD3/RD models, the KPZ 
stochastic growth equation is given by 

\begin{equation}  
\frac{\partial h({\mathbf x},t)}{\partial t}=F+\nu p^{2} \nabla ^{2}h({\mathbf
x},t)+  \frac {\lambda p^{3}}{2}[\nabla h({\mathbf x},t)]^{2}+\eta ({\mathbf
x},t).  \label{estocastica1} 
\end{equation}     

Also, for the case of LDS/RD, WV/RD, LC/RD and RDSR/RD models, 
the EW stochastic growth equation is given by

\begin{equation} 
\frac{\partial h({\mathbf x},t)}{\partial t}=F+\nu_{o} p^{2} \nabla ^{2}h({\mathbf x},t)+
\eta ({\mathbf x},t),
\label{estocastica2}
\end{equation} 

\noindent while for the case of the DST/RD model, the stochastic growth 
equation is given by

\begin{equation}\label{estocastica6}
\frac{\partial h}{\partial t} = -K p^{2} \nabla^{4} h(\textbf{r},t)+ 
\lambda_{1} p^{3} \nabla^{2}(\nabla h(\textbf{r},t))^{2} + 
\eta ( \textbf{r},t).
\end{equation}

Finally, in Table I we have summarized  the list of all possible
stochastic equations resulting for competitive deposition models belonging 
to four different universality classes when the competitive 
process can be mapped into the two types of random walks already 
considered.
 
\vspace{2cm}
{\bf Table I:} Summary of stochastic equations corresponding to models 
belonging to four different Universality Classes such that the
competitive process can be represented by random walks of Type A and B. 
More details in the text.\\ 

\vspace{0.3cm}
{\centering \begin{tabular}{|c|c|c|c|c|c|c|c|c|c|}
\hline 
Universality Class &
 Random Walk "A"&
 Random Walk "B" \\
\hline 
\hline 
Edwards-Wilkinson &
$\partial h({\mathbf x},t)/\partial t=\nu_{o} p \nabla 
^{2}h({\mathbf x},t)$&
$\partial h({\mathbf x},t)/\partial t=\nu_{o} 
p^{2} \nabla ^{2}h({\mathbf x},t)$
\\
   &
$+\eta ({\mathbf x},t)$&
$+ \eta ({\mathbf x},t)$
\\

\hline 
Kardar-Parisi-Zhang&
$\partial h({\mathbf x},t)/\partial t=\nu p \nabla ^{2}h({\mathbf
x},t)$&
$\partial h({\mathbf x},t)/\partial t=\nu p^{2} \nabla ^{2}h({\mathbf
x},t)$\\
&
$+  \lambda p^{3/2}[\nabla h({\mathbf x},t)]^{2}+\eta ({\mathbf
x},t)$&
$+  \lambda p^{3}[\nabla h({\mathbf x},t)]^{2}+\eta ({\mathbf
x},t)$\\

\hline 
Linear-MBE&
$\partial h({\mathbf x},t)/\partial t = -K p \nabla^{4} 
h({\mathbf x},t)$
&
$\partial h({\mathbf x},t)/\partial t = -K p^{2} \nabla^{4} 
h({\mathbf x},t)$
\\
 &
$ + \eta ({\mathbf x},t)$
&
$ + \eta ({\mathbf x},t)$
\\
\hline 
Non linear-MBE&
$\partial h({\mathbf x},t)/\partial t = -K p \nabla^{4} h({\mathbf x},t)$
&
$\partial h({\mathbf x},t)/\partial t = -K p^{2} \nabla^{4} h({\mathbf x},t)$
\\
&
$+ \lambda_{1} p^{3/2} 
\nabla^{2}(\nabla h({\mathbf x},t))^{2} + \eta ({\mathbf x},t)$
&
$+ \lambda_{1} p^{3} 
\nabla^{2}(\nabla h({\mathbf x},t))^{2} + \eta ({\mathbf x},t)$
\\

\hline 
\end{tabular}\par}
\vspace{3cm}

{\bf IV. CONCLUSIONS.}\\ 
 
We have studied a wide family of competitive growth models
(generically called X/RD models) 
where particles of the same type are aggregated 
according to the rules of a generic discrete model X with 
probability $p$ and according to the rules of 
random deposition (RD) with probability ($1-p$). 

First, we have focussed our study on the properties of the 
exponents related to the interface width 
$W_{sat} \propto p^{-\delta}$ and the characteristic 
crossover time $t_{x2} \propto p ^{-y}$.
We have shown that both exponents 
are not independent and one has that the exact 
relationship $\delta = y \beta_{RD}$
($\beta_{RD} = 1/2$) holds for the BD/RD and RDSR/RD models.
However, we expect that the above relationship would hold for 
any competitive growth model of the type X/RD.

Also, we have found that the values of the exponent 
$\delta$ do not significantly depend on the finite size of the sample. 
This property has allowed us to systematically study competitive 
growth models using lattices of modest size.  This study shows that $\delta$ 
exhibits universality and its values, in the limit $p \rightarrow 0$,
are restricted to either $\delta = 1/2$ or $\delta = 1$, depending on the 
model considered. Furthermore, by using a correspondence between 
two neighbouring sites in the discrete model $((h(i)-h(i+1)))$ and 
two types of random walks, we have determined the exact values of the 
exponent $\delta$. When the height difference between two neighbour 
sites corresponds to a random walk of Type A that in each cycle 
walks $\langle n \rangle$ steps and after that 
returns to a point at a distance proportional to its initial position,
one has $\delta=1/2$ and consequently $y=1$. 
On the other hand, when the height difference between 
two neighbouring sites corresponds to a random walk of Type B
that after $\langle n \rangle$ steps moves towards 
its initial position $l$ steps, we have found that $\delta = 1$ 
and $y = 2$.

Finally, using the exact values for the exponents $\delta$ and $y$, as
well as a scaling argument on $p$, we have derived the stochastic 
growth equations for the whole family of competitive models studied. So, 
we conclude that the properties of the height difference at 
saturation $(h(i)-h(i+1))$ in the {\bf discrete model}, 
which determines the behaviour of $W_{sat}$,
have allowed us to quantitatively determine the 
exact dependence on $p$ of the {\bf coarse-grained} stochastic equations.
  
It is worth mentioning that the derivation of this type of coarse-grained 
stochastic equations, based on the growing rules of the corresponding
microscopic models, is an interesting challenge in the field of modern 
Statistical Physics. So, we expect that the relationships between microscopic
parameters and stochastic equations obtained and discussed in this work will 
stimulate studies of this field.\\ 

{\bf  ACKNOWLEDGEMENTS}. This work was financially supported by 
CONICET, UNLP and  ANPCyT (Argentina).

\end{document}